\documentclass[11pt]{article}
\usepackage{amssymb}
\usepackage[colorlinks]{hyperref}
\usepackage[dvips]{graphicx,color}

\newcommand{\ket}[1]{\left | #1 \right \rangle}
\newcommand{\bra}[1]{\left \langle #1 \right |}

\def\openone{\leavevmode\hbox{\small1\kern-3.8pt\normalsize1}}

\def\cc{{\cal C}}
\def\cp{{\cal P}}

\newtheorem{theorem}{Theorem}

\newcommand{\beq}{\begin{equation}}
\newcommand{\eeq}{\end{equation}}
\newcommand{\beqa}{\begin{eqnarray}}
\newcommand{\eeqa}{\end{eqnarray}}

\begin{document}
\begin{center}
{\LARGE\bf Matchgates and classical simulation of \\[2mm] quantum circuits }\\
\bigskip
{\normalsize Richard Jozsa$^1$ and Akimasa Miyake$^{2,3}$}\\
\bigskip
{\small\it $^1$Department of Computer Science, University of
Bristol,\\ Merchant Venturers Building, Bristol BS8 1UB U.K.\\[1mm]
$^2$Institute for Theoretical Physics, University of Innsbruck,\\
Technikerstrasse 25, A-6020 Innsbruck, Austria.\\[1mm]
$^3$Institute for Quantum Optics and Quantum Information, \\
Austrian Academy of Sciences, Innsbruck, Austria.} \\[3mm]
\end{center}

\begin{abstract} Let $G(A,B)$ denote the 2-qubit gate which acts
as the 1-qubit $SU(2)$ gates $A$ and $B$ in the even and odd
parity subspaces respectively, of two qubits. Using a Clifford
algebra formalism we show that arbitrary uniform families of
circuits of these gates, restricted to act only on nearest
neighbour (n.n.) qubit lines, can be classically efficiently
simulated. This reproduces a result originally proved by Valiant
using his matchgate formalism, and subsequently related by others
to free fermionic physics. We further show that if the n.n.
condition is slightly relaxed, to allowing the same gates to act
only on n.n. and next-n.n. qubit lines, then the resulting
circuits can efficiently perform universal quantum computation.
From this point of view, the gap between efficient classical and
quantum computational power is bridged by a very modest use of a
seemingly innocuous resource (qubit swapping).  We also extend the
simulation result above in various ways. In particular, by
exploiting properties of Clifford operations in conjunction with
the Jordan-Wigner representation of a Clifford algebra, we show
how one may generalise the simulation result above to provide
further classes of classically efficiently simulatable quantum
circuits, which we call Gaussian quantum circuits.

\end{abstract}

\bigskip

\section{Introduction}\label{intro}

Quantum computation is widely regarded as being more powerful than
classical computation. Indeed in some scenarios there are provable
benefits, such as an exponential reduction in communication
resources for some distributed computing tasks (e.g. Raz 1999) and
in quantum cryptography, the ability to communicate with
unconditional security against eavesdropping. These results 
depend neither on any computational hardness assumptions nor on the
presence of any oracle relativisations. Furthermore in suitably
relativised oracle models of computation there are various known
exponential savings in quantum versus classical query complexity
such as Deutsch and Jozsa 1992 and Simon 1997. However for pure
(unrelativised) computation there is to date no proof of
separation and it is still possible that efficient classical and
quantum computational power might coincide i.e. the complexity
classes BPP and BQP might be equal. (See for example Nielsen and
Chuang 2000 for a definition of these classes. Here and below the
term ``efficient'' is synonymous with ``polynomial time''). Note
that this is not the same question as the issue of efficient
classical simulation of quantum processes since any BQP algorithm
is a quantum process of only a severely restricted kind, required
to satisfy infinitely many constraints relative to an infinite set
of input states viz. all computational basis states. In this paper
we will study a representation of quantum computation in which the
gap (if it exists) between efficient classical and efficient
quantum computation appears to be surprisingly fragile, being
provably bridged by a very modest use of a seemingly trivial
resource (cf theorems \ref{one} and \ref{two} below).

We will provide a self contained development of a class of quantum
circuits based on so-called matchgates, a notion that was
introduced in Valiant 2002. Our approach is closely related to
work of Knill 2001,  Terhal and DiVincenzo 2002 and DiVincenzo and
Terhal 2005, relating matchgate circuits to free fermionic quantum
computation (and later further extended in Bravyi 2005, 2008).
Here we will emphasise the underlying mathematical ingredients and
consider some further properties and generalisations that go
beyond the fermionic formalism. Indeed the existence of a physical
interpretation in terms of fermionic physics, although
interesting, appears to be entirely fortuitous and of no
particular consequence for our considerations of computational
complexity issues {\em per se}.

Consider 2-qubit gates $G(A,B)$ of the form (in the computational
basis):
\begin{equation}\label{gab} G(A,B) = \left(
\begin{array}{cccc} p&0&0&q \\ 0&w&x&0 \\ 0&y&z&0 \\ r&0&0&s
\end{array} \right) \hspace{1cm} A = \left( \begin{array}{cc}
p&q \\ r&s \end{array} \right) \hspace{5mm} B= \left(
\begin{array}{cc} w&x \\ y&z \end{array} \right) \end{equation}
where $A$ and $B$ are both in $SU(2)$ or both in $U(2)$ with the
{\em same determinant}. Thus the action of $G(A,B)$  amounts to
$A$ acting in the even parity subspace (spanned by $\ket{00}$ and
$\ket{11}$) and $B$ acting in the odd parity subspace (spanned by
$\ket{01}$ and $\ket{10}$). Occasionally we will wish to consider
2-qubit gates of the form eq. (\ref{gab}) but having det $A \neq$
det $B$. In this case the gate will be denoted $\tilde{G}(A,B)$.
To emphasise this distinction we sometimes refer to a gate with
det $A =$ det $B$ as an {\em allowable} $G(A,B)$ gate. We will
denote the Pauli operators by $X$, $Y$, and $Z$.

\begin{theorem}\label{one} Consider any uniform (hence poly-sized)
quantum circuit family comprising only $G(A,B)$ gates such that:\\
(i)
the $G(A,B)$ gates act on nearest neighbour (n.n.) lines only;\\
(ii) the input state is any product state;\\ (iii) the output is a
final measurement in the computational basis on any single line.\\
Then the output may be classically efficiently simulated. More
precisely for any $k$ we can classically efficiently compute the
expectation value $\langle Z_k\rangle_{\rm out} = \bra{\psi_{\rm
out}} Z_k \ket{\psi_{\rm out}} = p_0-p_1$ where $Z_k$ is the Pauli
$Z$ operator on the $k^{\rm th}$ line, $\ket{\psi_{\rm out}}$ is
the final state and $p_0,p_1$ are the outcome probabilities.
\end{theorem}

Theorem \ref{one} is very similar to the classical simulation
result of Valiant 2002 and Terhal and DiVincenzo 2002. Our result
is more general in the feature of allowing arbitrary product state
inputs (rather than just computational basis states).  It is more
restrictive in considering only single bit outputs (rather than
individual probabilities of computational basis measurements
across many lines) and in not encompassing the adaptive circuits
that are included in Terhal and DiVincenzo 2002.

The notion of {\em efficient classical simulation} that we will
use in this paper is the following. Let $C_n$ be any uniform
family of quantum circuits together with (i) a specified class of
input states (usually taken to be product states, but we may also
restrict to just computational basis states) and (ii) a specified
class of output measurements (which we take to be $Z_k$, a
$Z$-basis measurement on any single line). We say that $C_n$ is
{\em classically efficiently simulatable} (relative to (i) and
(ii)) if the probabilities of measurement outcomes can be {\em
computed} by classical means to $m$ digits of accuracy in
poly$(n,m)$ time.

Note that this ability to efficiently {\em compute} the
probabilities to {\em exponential} accuracy is a rather strong
notion e.g. we might instead adopt weaker criteria such as the
ability to compute the probabilities to accuracy $1/{\rm poly}(m)$
i.e. $O( \log m)$ digits in poly$(n,m)$ time, or the ability to
{\em sample} the output probability distribution {\em once} (by
classical efficient probabilistic means, to suitably accuracy).
Indeed the last requirement would suffice in issues of the
comparison of quantum to classical computational power but we will
in fact achieve the strongest notion above in our results and we
thus adopt it as our definition. We make further comments about
implications of this strong notion of classical simulation in the
concluding section \ref{concls} below.

Note that if the n.n. $G(A,B)$ circuits in theorem \ref{one} are
considered with computational basis inputs and also required to
satisfy the BQP bounded probability conditions (viz. that the
output probabilities are always $\geq\frac{2}{3}$ or
$\leq\frac{1}{3}$), then the ability to classically {\em
calculate} the output probabilities (rather than the ability
merely to sample the output distribution once) implies that the
corresponding decision problem is not just in BPP but actually in
P i.e. {\em deterministic} classical polynomial time.

In the following sections we will first prove a universality
result for $G(A,B)$ gates, if these gates are also allowed to act
on next-n.n. lines in addition to the n.n. lines of theorem
\ref{one}. Then we give some background on the origin of the
notion of matchgates, which first lead to the consideration of
circuits of $G(A,B)$ gates in Valiant 2002. Next we consider a
formalism of anti-commuting variables that form a Clifford
algebra, leading to a proof of theorem \ref{one}. This approach is
essentially the one given in Knill 2001 and Terhal and DiVincenzo
2002, but we give some more transparent proofs and we develop
further properties. First, we elucidate the n.n. condition (i) in
theorem \ref{one}: we show that the general class of simulatable
gates comprises a uniformly describable family that may act on any
number of the $n$ qubits, in which the $G(A,B)$ gates appear as
the subset of n.n. 2-qubit gates. We show furthermore that all
gates in the family can be obtained as {\em circuits} of n.n.
$G(A,B)$ gates (hence adding nothing new) and that non-n.n.
$G(A,B)$ gates are not generally in the family. Second, by
considering Clifford {\em operations}\footnote{The appellation
``Clifford'' here, commonly used in quantum computation
literature, appears not to be mathematically related to the well
established notion of Clifford {\em algebra} in mathematics
generally.}  (i.e. $n$-qubit unitary operations that normalise the
$n$-qubit Pauli group in $U(2^n)$) in conjunction with the
Clifford algebra formalism, we will describe an avenue for
generalising theorem \ref{one} and give some examples of
simulatable circuits which
 cannot be obtained as circuits of n.n. $G(A,B)$ gates only.
Since all our classes of classically simulatable circuits comprise
gates that are generated by Hamiltonians expressible as quadratic
elements of a Clifford algebra, we call them {\em Gaussian quantum
circuits}.

\section{Universality of n.n. and next-n.n. $G(A,B)$ gates}\label{nnngates}

The n.n. condition in theorem \ref{one}(i) is perhaps a surprising
ingredient but it is crucial: it was already mentioned in Terhal
and DiVincenzo 2002 (based on a result in Kempe {\em et al.}
2001) that n.n. $G(A,B)$ gates together with the swap gate $SWAP$,
or equivalently $G(A,B)$ gates acting on arbitrary pairs of qubit
lines,  can perform universal quantum computation. We will prove a
stronger result:

\begin{theorem}\label{two} Let $C_n$ be any uniform family of
quantum circuits with output given by a $Z$ basis measurement on
the first line. Then $C_n$ may be simulated by a circuit of
$G(A,B)$ gates acting on n.n. or next n.n. lines only (i.e. on
line pairs at most distance 2 apart) with at most a constant
increase in the size of the circuit.
\end{theorem}

Before the proof we make a few remarks. As an immediate corollary
we have that any BQP algorithm can be simulated by a poly-sized
circuit of $G(A,B)$ gates acting only on n.n. and next-n.n. lines.
This fact together with theorem \ref{one} shows that a very
limited use of the seemingly innocuous operation $SWAP$ (on n.n.
lines) allowing n.n. $G(A,B)$ gates to act on lines just one
further apart, suffices to bridge the gap between classical and
quantum efficient computational power. The result becomes perhaps
even more striking if we note that $SWAP$ itself is very close to
being expressible in the allowed $G(A,B)$ form. Indeed $SWAP =
\tilde{G}(I,X)$ and fails only through a mere minus sign in ${\rm
det}\,X = - {\rm det}\,I$. Thus if we drop the ${\rm det}A= {\rm
det}B$ condition in eq. (\ref{gab}), then the resulting
$\tilde{G}(A,B)$ gates acting on n.n. lines become efficiently
universal for quantum computation.

The significance of $SWAP$ (or equivalently the ability of 2-qubit
gates to act on distant lines) for quantum computational power
appears also in a different context. Using the formalism of tensor
network contractions it may be shown (Markov and Shi 2008, Jozsa
2006 and Yoran and Short 2006) that any poly-sized quantum circuit
of 1- and 2- qubit gates, which has log depth and in which the
2-qubit gates are restricted to act at bounded range (i.e. on line
pairs at most distance $c$ apart, for some constant $c$) may be
classically efficiently simulated. It is also known (Cleve and
Watrous 2000) that Shor's quantum factoring algorithm (Shor 1997)
can be implemented as a log depth circuit but its 2-qubit gates
act on distant lines, $O(n)$ apart, which is not bounded with
increasing input size $n$. Thus from this point of view, the
quantum advantage of the algorithm (over classical algorithms)
rests entirely on the presence of {\em unboundedly} distant
actions (or unbounded use of $SWAP$). Also it is shown in Terhal
and DiVincenzo 2004 and Jozsa 2006 that all depth 2 circuits
(followed by a measurement) are classically efficiently
simulatable even if 2-qubit gates act on arbitrary line pairs
while the same simulation result for depth 3 circuits (with a
suitably strong notion of classical simulation) would imply
equality of BPP and BQP. Here again the feature of unboundedly
distant action is essential, whereas our result in theorems
\ref{one} and \ref{two} achieves full efficient quantum
computational power by passage from distance one to just bounded
distance two.

\begin{figure}[t]
\begin{center}
\includegraphics[width=0.8\textwidth,clip]{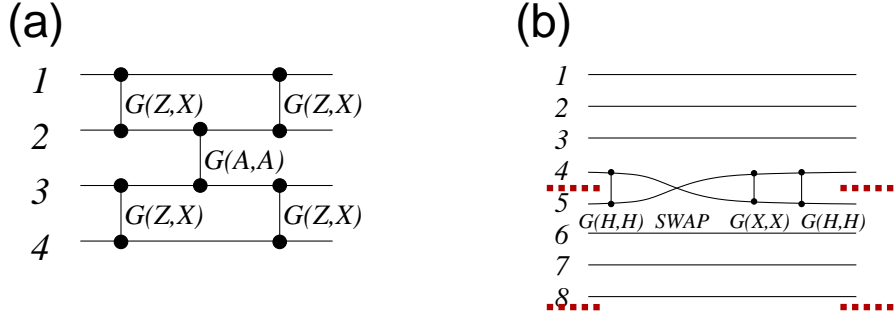}
\end{center}
\caption{Encoded universality by n.n. and next n.n. $G(A,B)$
gates. The logical single-qubit unitary gate $A$ and the logical
two-qubit $CZ$ gate are illustrated in (a) and (b), respectively.}
\label{fig:universal_valiant}
\end{figure}

\noindent {\bf Proof of theorem \ref{two}}:\,\, Given any uniform
quantum circuit family we may assume w.l.o.g. that it comprises
n.n. controlled-$Z$ gates (n.n. $CZ$) and 1-qubit gates generically
denoted as $A$.

We start with a {\em quadrupled} number of qubit lines and encode
the original input $\ket{0}$'s and $\ket{1}$'s as logical basis
states $\ket{0_L}=\ket{0000}$ and $\ket{1_L}=\ket{1001}$
respectively in consecutive blocks of 4 lines each (cf the remark
after the proof). Then with suitably encoded gate operations the
whole computation will stay within tensor products of span$\{
\ket{0_L},\ket{1_L}\}$ of each quadruple of qubits.

On any such quadruple of lines, say 1234, we can perform the
encoded 1-qubit gate $A$ as the following sequence of allowed n.n.
gates (depicted in figure \ref{fig:universal_valiant} (a)):
\begin{equation}\label{gaa} G(Z,X)_{12}\, G(Z,X)_{34}\,
G(A,A)_{23}\, G(Z,X)_{12}\, G(Z,X)_{34}
\end{equation}
(where the subscripts denote the line numbers). To see this, note
that
\[ G(Z,X)=\tilde{G}(Z,I)\tilde{G}(I,X)=(CZ)(SWAP) ,\]
so that $G(Z,X)_{12}G(Z,X)_{34}$ can be thought of (for our
logical basis states) as just swapping lines 1 and 4 into
positions 2 and 3. In view of the form of the encoding $\ket{0_L}$
and $\ket{1_L}$, the logical qubit is then encoded in the $\{
\ket{00},\ket{11} \}$ subspace of lines 2 and 3 so $G(A,A)$ will
apply the 1-qubit gate $A$ to it. Finally the lines are swapped
back to their original positions, restoring the encoding.

To perform an encoded $CZ$ on two consecutive quadruples, say 1234
and 5678 we simply apply $CZ_{45}$ i.e. $CZ$ on the ``crossover''
pair of lines 4 and 5. Indeed for any pair of basis states
$\ket{abcd}_{1234}$ and $\ket{efgh}_{5678}$ we'll get a minus sign
iff $d=e=1$, giving the correct action on any encoded $\ket{x_L}$
and $\ket{y_L}$. Next note that since composition of $G(A,B)$
gates amounts to multiplying the $A$'s and $B$'s separately we
obtain (with all gates acting on lines 4 and 5)
\begin{equation}\label{cz45}
CZ_{45}=\tilde{G}(Z,I)=G(H,H)\tilde{G}(X,I)G(H,H)=
G(H,H)G(X,X)\tilde{G}(I,X)G(H,H)
\end{equation}
where $H = \frac{1}{\sqrt{2}}(X + Z)$ is the Hadamard operator. In
the last expression, $\tilde{G}(I,X)$ is $SWAP$ and all other
gates are allowable $G(A,B)$ gates. This implementation of
$CZ_{45}$ is depicted in figure \ref{fig:universal_valiant} (b).

Finally note that in an arbitrary such circuit, $SWAP$ is used
only on ``crossover'' pairs (4,5), (8,9) etc. of the encoding
quadruples (1,2,3,4), (5,6,7,8), (9,10,11,12) etc. Hence no line
is ever moved more than one position distant from its original
location, by overall action of any number of such $SWAP$s. We may
commute all these $SWAP$ operations out to the output end of the
circuit. In so doing, any line of each (originally n.n.) $G(A,B)$
gate in eq. (\ref{gaa}) may be moved by at most one place but we
can never move both lines of any n.n. $G(A,B)$ gate in view of the
block size 4 of the encoding. Thus the resulting circuit (with
$SWAP$'s eliminated) comprises only $G(A,B)$ gates acting on n.n.
or next-n.n. lines, as required.

In this process $SWAP$ gates need to be commuted across $G(X,X)$
and $G(H,H)$ gates (cf eq. (\ref{cz45})). But for any $G(A,B)$ we
have (using $SWAP=\tilde{G}(I,X)$) that
\[ G(A,B) \,\, SWAP=SWAP\,\, (SWAP\,\, G(A,B)\,\, SWAP)= SWAP\,\, G(A,XBX) \]
and the last gate is an allowed $G(A,B)$ gate. Since the whole
computation is engineered to represent the original given circuit,
recoded in the $\{ \ket{0_L}=\ket{0000},\ket{1_L}=\ket{1001} \}$
subspaces of consecutive line quadruples, a final measurement on
line 1 will produce the same output distribution as a measurement
on qubit 1 in the original given circuit. This completes the proof
of theorem \ref{two}. $\Box$

We remark that instead of the quadruple encoding above, we might
have considered the simpler $\ket{0_L}=\ket{00}$ and $\ket{
1_L}=\ket{11}$ as a potentially more natural choice. Indeed in
that case the 1-qubit gate $A$ is applied very simply as $G(A,A)$
(in contrast to eq. (\ref{gaa})) but the $SWAP$'s from the $CZ$
actions may now move both lines of a n.n. $G(A,B)$ gate in
opposite directions, resulting in $G(A,B)$'s on lines up to
distance 3 (rather than just 2) apart.

\section{Perfect matchings and matchgates}\label{matchmatch}

Before beginning our development of theorem \ref{one} we give some
brief background remarks on the interesting provenance of
Valiant's notion of matchgates. (These remarks will not be used in
any further results). Matchgates arose (Valiant 2002, 2007) in the
context of the theory of perfect matchings in graphs. For a graph
$G$ a perfect matching is a set $M$ of edges such that each vertex
is the endpoint of exactly one edge in $M$. It is known that the
problem of counting the number of perfect matchings in a graph is
computationally very hard  (being complete for the complexity
class \#P, c.f. Papadimitriou 1994) but for {\em planar} graphs it
is, remarkably, computable in polynomial time, using the
Fisher-Kasteleyn-Temperley (FKT) algorithm (Kasteleyn 1961,
Temperley and Fisher 1961 and Jerrum 2003).

More generally we may consider weighted graphs $G$ in which each
edge $(ij)$ is assigned a weight $w_{ij}$ and introduce the
so-called match sum:
\begin{equation}\label{pmg} 
PerfM\,(G)=\sum_{\begin{array}{c} \mbox{\small perfect matchings} 
\\ \mbox{\it \small M} \end{array}}  \prod_{(ij) \in M} w_{ij} .
\end{equation} Then the FKT algorithm provides a polynomial time
computation of the match sum for any planar graph.

Next consider a (planar) weighted graph with a designated set $\{
v_1, \ldots ,v_n \}$ of ``input'' vertices and a disjoint
designated set $\{ v_1', \ldots ,v_m' \}$ of ``output'' vertices,
and consider the indexed collection (tensor)
\[ M^{i_1 \ldots i_n}_{j_1 \ldots j_m} = PerfM
(G^{i_1 \ldots i_n}_{j_1 \ldots j_m}) \] where each index takes
values 0 and 1 and $G^{i_1 \ldots i_n}_{j_1 \ldots j_m}$ is the
graph obtained from $G$ by deleting all those input and output
vertices (together with their incident edges) that have
corresponding index $i$ or $j$ set to 1. Hence the tensor
components are each computable in poly time by the FKT algorithm.

Matchgates are essentially these tensors with some additional
technical modifications (Valiant 2002, 2007) whose specification
we will omit here. Suffice it to say that the full definition is
chosen so that a circuit of matchgates, representing contraction
of a matchgate tensor network, corresponds to the problem of
evaluating the match sum of $G_{\rm tot}$  where $G_{\rm tot}$
(with possibly some residual uncontracted input and output
vertices) is the graph obtained from the graphs of the individual
matchgates by identifying (or ``glueing along'') input and output
vertices that are contracted in the tensor network. It follows
that the components of the contraction are computable in poly time
too. This is essentially the content of Valiant's so-called Holant
theorem (Valiant 2007). The expression and clarification of the
Holant theorem in terms of tensor contractions (and its invariance
under appropriate basis changes in representing the tensors) was
developed in a series of works by Cai and Choudhary 2006a,b.

For some choices of graphs and weights (with equal numbers $m=n$
of input and output nodes) the matchgate tensors can be {\em
unitary} i.e. unitary operations on $m$ qubits. Hence the above
formalism leads to a class of quantum circuits (comprising unitary
matchgates) that can be classically efficiently simulated. For
example it may be shown (Valiant 2002) that the unitary gates
$G(A,B)$ in eq. (\ref{gab}) arise as matchgates with 2 input and 2
output vertices.

The FKT algorithm (Kasteleyn 1961, Temperley and Fisher 1961 and
Jerrum 2003) proceeds by setting up a suitable antisymmetric
incidence matrix $A$ of the graph's weights and then computing
$PerfM (G)$ as the Pfaffian of $A$ (which in fact equals
$\sqrt{{\rm det} A}$). Like the determinant of an arbitrary
matrix, the definition of Pfaffian of an antisymmetric matrix is
an expression involving exponentially many terms a priori yet
computable on polynomial time. It is known that Pfaffians also
occur in the mathematical formalism of fermionic quantum physics
which suggests that there may be some relationship (or at least
some form of translation of basic problems) between fermionic
physics and perfect matchings in graphs. Indeed soon after the
appearance of Valiant's work (Valiant 2002) on classical
simulation of matchgate quantum circuits, Knill 2001 and Terhal
and DiVincenzo 2002 provided an interpretation of it in terms of
fermionic quantum gates and this formalism was subsequently
further developed by Bravyi 2005, 2008.

\section{Clifford algebras, quadratic Hamiltonians and classical
simulation}\label{cliffquad}

We now return to developing a formalism for treating theorem \ref{one}
and some generalisations.
For $n$ qubit lines, we introduce the set of $2n$ hermitian
operators $c_\mu$ which satisfy the anti-commutation relations,
\begin{equation}\label{cliffcomm}
\{ c_\mu, c_\nu \} \equiv c_\mu c_\nu + c_\nu c_\mu =
2 \delta_{\mu\nu} I \hspace{1cm} \mu,\nu = 1, \ldots ,2n.
\end{equation}
These relations define a {\em Clifford algebra} $\cc_{2n}$ on $2n$
generators whose elements are arbitrary complex linear
combinations of products of generators \footnote{As mentioned in
Knill 2001 and Somma {\em et al.} 2006, it is possible to consider
$2n+1$ anti-commuting operators to define the Clifford algebra and
correspondingly to have $SO(2n+1)$ symmetry in theorem \ref{quadh}
below. However, this extension appears not to lead to a
significant generalization of our results.}. Since each generator
squares to the identity a general element in the algebra may be
expressed as a polynomial of degree at most $2 n$,
\begin{equation}\label{cliffelt} \sum_{i_1 < \ldots < i_k} A_{i_1
\ldots i_k} c_{i_1} \ldots c_{i_k} \end{equation} (where the index
set $\{ i_1, \ldots , i_k \}$ may be empty). It follows from eq.
(\ref{cliffcomm}) that the monomials in the sum are linearly
independent so as a vector space $\cc_{2n}$ has dimension
$2^{2n}=2^n \times 2^n$. Hence (hermitian) matrix representations
of the $c_\mu$'s will involve matrices of size $2^n \times 2^n$.

Operators $c_\mu$ satisfying eq. (\ref{cliffcomm}) arise in
the formalism of fermionic physics where they are known as
Majorana spinors. In that formalism we start with a set of
operators $a_1, \ldots , a_n$ associated to $n$ free fermionic
modes, satisfying the standard anti-commutation relations for
fermionic creation and annihilation operators:
\[
\{ a_i,a_j\} \equiv a_ia_j+a_ja_i =0=\{ a_i^\dagger, a_j^\dagger
\} \hspace{1cm} \{ a_i,a_j^\dagger \} = \delta_{ij}I .
\]
Then as a consequence of these relations, the following hermitian
operators (which are fermionic analogues of position and momentum
operators):
\[
c_{2k-1}=a_k+a_k^\dagger \hspace{1cm} c_{2k}= (a_k-a_k^\dagger)/i
\hspace{1cm} k=1,\ldots,n ,
\] satisfy the Clifford algebra relations eq. (\ref{cliffcomm}).
However we emphasise that in the present paper we are not
concerned with the study of free fermions {\em per se} but rather,
consideration of general quantum circuit simulation properties,
based on the Clifford algebra structure, which can also go beyond
the fermionic formalism (such as the statement in theorem
\ref{two} and results in section \ref{clops} below).

For theorem \ref{one} and generalisations we will (in later
sections) consider matrix representations of the Clifford algebra
but we first develop some further abstract algebra. A {\em
quadratic Hamiltonian} is an element of $\cc_{2n}$ of the form,
\begin{equation}
H = i\sum_{\mu \neq \nu = 1}^{2n}
h_{\mu\nu} c_\mu c_\nu
\label{quad_h}
\end{equation}
where $h_{\mu\nu}$ is a $2n \times 2n$ matrix of coefficients.
Note that we omit $\mu=\nu$ terms which contribute only an overall
additive constant to $H$. Since $c_\mu c_\nu=-c_\nu c_\mu$ and
imposing $H=H^\dagger$ we may w.l.o.g.
take $h_{\mu\nu}$ to be a real antisymmetric matrix.

Given $H$ we consider the unitary operation $U=e^{iH}$ (where the
exponential is calculated in the algebra of $\cc_{2n}$ as the
power series). Any such unitary operation corresponding to a {\em
quadratic} Hamiltonian is called a {\em Gaussian} operation. The
following result from fermionic linear optics (cf Knill 2001,
Terhal and DiVincenzo 2002, DiVincenzo and Terhal 2005, Bravyi
2005) will be basic for our classical simulation results and we
include a simple proof of it.

\begin{theorem}\label{quadh} Let $H$ be any quadratic Hamiltonian
and $U=e^{iH}$ the corresponding Gaussian operation. Then for all
$\mu$: \[ U^{\dagger} c_\mu U = \sum_{\nu=1}^{2n} R_{\mu\nu}c_\nu
\] where the matrix $R$ is in $SO(2n)$, and we obtain all of
$SO(2n)$ in this way. In fact $R=e^{4h}$.
\end{theorem}

\noindent {\bf Proof:}\,\, Write $c_\mu$ as $c_{\mu}(0)$
and introduce $c_{\mu} (t)=U(t) c_{\mu}(0) U(t)^\dagger$ with
$U(t)=e^{iHt}$. Then \[
\frac{d c_{\mu}(t)}{dt}=i[H, c_{\mu}(t)] \] (with square
brackets $[a,b]$ denoting the commutator $ab-ba$).  But
$[c_{\nu_1}c_{\nu_2}, c_\mu ]=0$ if $\mu \neq \nu_1,\nu_2$ and $[
c_\mu c_\nu,c_\mu]=-2c_\nu$ (using eq.(\ref{cliffcomm})) so
\[ \frac{d c_{\mu}(t)}{dt}= \sum_\nu 4 h_{\mu\nu} c_{\nu}(t)
\hspace{1cm} \mbox{and hence}\hspace{1cm} c_{\mu}(t)=\sum_\nu
R_{\mu\nu}(t) c_{\nu}(0) \] where $R=e^{4ht}$. It is well known
that antisymmetric matrices are the infinitesimal generators of
rotations and the theorem follows by just setting $t=1$.\,$\Box$

The significance of theorem \ref{quadh} for us is the following:
note that $e^{iH}$ generally involves all products of all
generators so the expression $U c_\mu U^\dagger$ could potentially
finish up anywhere in the exponentially large
($2^{2n}$-dimensional) linear space $\cc_{2n}$. However it always
happens to stay within the {\em polynomially small}
($2n$-dimensional) subspace spanned by just the generators
themselves. We exploit this feature of the adjoint representation
(cf. also Somma {\em et al.} 2006 for a more general
Lie-theoretic setting), using the following strategy.

We find a hermitian representation of the $c_\mu$'s on $n$-qubit
operators and then the Gaussian operations corresponding to
quadratic Hamiltonians define a class of $n$-qubit unitary gates.
Let $\tilde{U}$ be any circuit of these with $\ket{\psi_{\rm
out}}=\tilde{U}\ket{\psi_{\rm in}}$ for some choice of input
state. Then by theorem \ref{quadh} for each $\mu$ we have the
expectation value
\begin{equation}\label{sumc} \langle c_\mu \rangle_{\rm out}=
\bra{\psi_{\rm in}} \tilde{U}^\dagger c_\mu \tilde{U}\ket{\psi_{\rm in}}=
\sum_{\nu=1}^{2n} \tilde{R}_{\mu\nu} \bra{\psi_{\rm in}} c_\mu
\ket{\psi_{\rm in}} \end{equation} where $\tilde{R}_{\mu\nu}$ is the
product of all $SO(2n)$ matrices corresponding to the individual
gates of $M$. Hence the full matrix $\tilde{R}_{\mu\nu}$ is poly time
computable.

Now suppose further that $\ket{\psi_{\rm in}} = \ket{\xi_1}\ldots
\ket{\xi_n}$ is a {\em product} state and that $c_\mu$ is
represented by a {\em product} operator $P_1\otimes \ldots \otimes
P_n$. Then $\bra{\psi_{\rm in}} c_\mu \ket{\psi_{\rm in}} =
\prod_{i=1}^n \bra{\xi_i}P_i\ket{\xi_i}$ is also poly time
computable and hence $\langle c_\mu \rangle_{\rm out}$ is poly
time computable for each $\mu$.

However we really want $\langle Z_k\rangle_{\rm out} = p_0-p_1$
where $Z_k$ is the Pauli  $Z$ operator acting on the $k^{\rm th}$
line. Recall that $\cc_{2n}$ as a vector space has dimension
$2^n\times 2^n$ so it spans all $n$-qubit matrices in our
representation. Thus $Z_k$ must be expressible as some polynomial
of the form eq. (\ref{cliffelt}). If this polynomial has a
constant degree, independent of $n$, then $\langle Z_k
\rangle_{\rm out}$ will be poly time computable too. As an example
suppose $Z_1= -i c_1c_2$. Then

\begin{eqnarray}
\langle Z_1 \rangle_{\rm out} & = &
\bra{\psi_{\rm in}} (-i)\tilde{U}^\dagger c_1c_2 \tilde{U}
\ket{\psi_{\rm in}}  =
\bra{\psi_{\rm in}}(-i) (\tilde{U}^\dagger c_1 \tilde{U})
(\tilde{U}^\dagger c_2 \tilde{U})
 \ket{\psi_{\rm in}} \nonumber \\
  & = & \sum_{\nu_1 \ne \nu_2 =1}^{2n} \tilde{R}_{1\nu_1}
  \tilde{R}_{2\nu_2}\bra{\psi_{\rm in}} (-i)c_{\nu_1}c_{\nu_2}
  \ket{\psi_{\rm in}} .
  \label{sumz}
\end{eqnarray}

If the $c_\mu$ are product operators then so are all the monomials
such as $c_{\nu_1}c_{\nu_2}$ and $\bra{\psi_{\rm in}}
c_{\nu_1}c_{\nu_2}\ket{\psi_{\rm in}}$ will be poly time
computable for any product state input $\ket{\psi_{\rm in}}$. Note
also that the size of the sum in eq. (\ref{sumz}) is $O(n^2)$
(compared to the $O(n)$ sized sum for $\langle c_\mu \rangle_{\rm
out}$ in eq. (\ref{sumc}) and hence $\langle Z_1 \rangle_{\rm
out}$ is poly time computable too. This argument is easily
generalized to give the following result.

\begin{theorem}\label{simcost}
Consider any poly-sized circuit of Gaussian gates acting on a
product input state. If the observable $Z_k$ in the final
measurement is expressible in $\cc_{2n}$ as a polynomial of degree
$d$, then for each of its monomials the corresponding sum as in
eq. (\ref{sumz}) for $\langle Z_k\rangle_{\rm out}$ will be
$O(n^d)$ sized and hence $\langle Z_k\rangle_{\rm out}$ will be
poly time computable if $d$ does not increase with $n$.
\end{theorem}

\section{The Jordan-Wigner representation and theorem
\ref{one}}\label{jorwigone}

Introduce the $2n$ hermitian operators on $n$-qubits (omitting tensor
product symbols $\otimes$ throughout):
\begin{equation}\label{jwr} \begin{array}{cccccc}
c_1=X\,I\ldots I & & c_3= Z\,X\,I\ldots I & \quad\cdots\quad
& c_{2k-1}= Z\ldots Z\,X\,I\ldots I & \quad\cdots
\\
c_2=Y\,I\ldots I & & c_4= Z\,Y\,I\ldots I & \quad\cdots\quad &
\,\,c_{2k}\,\,\,\, = Z\ldots Z\,Y\,I\ldots I & \quad\cdots
\end{array}
\end{equation} where $X$ and $Y$ are in the $k^{\rm th}$ slot for
$c_{2k-1}$ and $c_{2k}$, and $k$ ranges from 1 to $n$. Thus the
operators $c_{2k-1},c_{2k}$ are associated to the $k^{\rm th}$
qubit line. It is straightforward to check that these matrices
satisfy the relations eq. (\ref{cliffcomm}) so we have a
representation of the Clifford algebra $\cc_{2n}$, known as the
Jordan-Wigner representation (Jordan and Wigner 1928). This is in
fact the unique representation of $\cc_{2n}$ up to a {\em global
unitary equivalence.} Furthermore, $Z_k=-i c_{2k-1}c_{2k}$, which
has bounded degree two, and the $c_\mu$'s are all product
operators. Hence for any poly sized circuit of Gaussian gates with
a product state input, $\langle Z_k \rangle_{\rm out}$ is poly
time computable. But what do these Gaussian gates actually look
like?

Consider first just qubit lines 1 and 2 (i.e. $c_1,c_2,c_3$ and
$c_4$) and corresponding quadratic Hamiltonians which involve 6
possible terms:
\[ \begin{array}{ccc} -ic_1c_2=ZI & & -ic_2c_3=XX \\ ic_1c_3=YX & &
-ic_2c_4=XY \\ ic_1c_4=YY & & -ic_3c_4=IZ. \end{array} \] These
operators are all trace free and all preserve the even and odd
parity subspaces. Hence the corresponding 6 parameter family of
Gaussian gates must be $SU(2)\oplus SU(2)$ decomposed relative to
the two parity subspaces. More explicitly we may first construct
the Pauli $X,Y,Z$ operators acting within the two subspaces (e.g.
$\frac{1}{2}(XX+YY)$ is $X$ acting in the odd subspace relative to
the $\{ \ket{01}, \ket{10} \}$ basis, and maps the even subspace
to zero) and generate the two $SU(2)$'s by direct exponentiation.
Hence we get precisely the $G(A,B)$ gates for lines 1 and 2 as the
Gaussian operations $U=e^{iH}$ with the quadratic Hamiltonian of
eq.~(\ref{quad_h}) restricted to use of $c_1, c_2, c_3, c_4$ only.
Similarly for any pair of {\em consecutive} lines we get all n.n.
$G(A,B)$ gates (since for lines $k,k+1$ the initial $Z$ operators
in eq. (\ref{jwr}) are eliminated in all quadratic products $c_\mu
c_\nu$ and the calculation proceeds exactly as above).

Thus all n.n. $G(A,B)$ gates are Gaussian for the Jordan-Wigner
representation and this completes the proof of theorem \ref{one}.

But there are still more Gaussian gates, generated by quadratic
Hamiltonians involving more $c_\mu$'s associated to a larger
number and more distant lines. Note first that if we use only the
four $c_\mu$'s associated to a pair of not-n.n. lines, we do {\em
not} get the corresponding (now non-n.n.) $G(A,B)$ gate acting on
those lines.  For example consider the quadratic term $c_2c_4$
(associated to n.n. lines 1 and 2) replaced by $c_2c_6$ (being the
corresponding operator associated to lines 1 and 3). We have
$c_2c_4=X_1Y_2$ but $c_2c_6= X_1Z_2Y_3$. Hence exponentiation of
the latter does not correspond to exponentiation of XY for lines 1
and 3 but gives a gate acting nontrivially across all three lines.

In summary so far, we see that non-n.n. $G(A,B)$'s are not
generally Gaussian. But n.n. $G(A,B)$'s are all examples of
Gaussian operations, albeit only special cases in the full set of
such operations that may generally act on any number of qubit
lines. Finally we show that these apparently more general Gaussian
operations actually bring nothing new, in the context of {\em
circuits} of gates:
\begin{theorem}\label{gausscircuit} Let $H=i\sum_{\mu,\nu}
h_{\mu\nu} c_\mu c_\nu$ be any quadratic Hamiltonian with
corresponding Gaussian gate $V=e^{iH}$ on $n$ qubits. Then $V$ as
an operator on $n$ qubits is expressible as a circuit of $O(n^3)$
n.n. $G(A,B)$ gates i.e. $V=U_N U_{N-1}\ldots U_1$ where each
$U_j=e^{iH_j}$ having $H_j=i\sum_{\mu,\nu} h_{\mu\nu} c_\mu c_\nu$
with the sum involving only four $c$'s associated to two n.n.
lines viz. $c_{2k-1},c_{2k},c_{2k+1},c_{2k+2}$ for some fixed $k$.
\end{theorem}

Note that (as shown in the proof below) the circuit expression
of theorem~\ref{gausscircuit} is {\em exact, analytic, and
explicitly describable in poly-time}, in contrast to an
alternative standard, but generally inefficient, asymptotic
decomposition utilising the Lie-Trotter expansion (for an
exponential of a sum of generally non-commuting operators).

\noindent {\bf Proof:}\,\, Let $V=e^{iH}$ be any Gaussian
operation as above. We have
\[
V^\dagger c_\mu V=\sum_{\nu =1}^{2n} R_{\mu\nu}c_\nu
\]
with $R\in SO(2n)$. We can efficiently decompose $R$ into its
generalized Euler angles (by the algorithm of section 4 in Hoffman
{\em et al.} 1972), obtaining $R=r_1 r_2\ldots r_M$ where
$M=O(n^2)$ and each $r_j$ is a rotation in $2n$ dimensions that
acts non-trivially only in 2-dimensions, spanned by say the
$a^{\rm th}$ and $b^{\rm th}$ co-ordinates. Thus $r_j =e^{4h_j}$
where $h_j$ is an antisymmetric matrix with nonzero values
(denoted $\pm \theta/2$) only in its $a^{\rm th}$ and $b^{\rm th}$
columns and rows. Then introduce $H_j=i \theta c_{a} c_{b}$ so
$U_j=e^{iH_j}$ has
\[
U_j^\dagger c_\mu U_j=\sum_{\nu =1}^{2n} (r_j)_{\mu\nu}c_\nu .
\]
In this construction $c_a$ and $c_b$ do not generally belong to
n.n. qubit lines. To remedy this we introduce the n.n. ``modified
swap'' operation (Bravyi and Kitaev 2002) defined, for example for
lines 1 and 2, by
\begin{equation}\label{s12}
S_{12} = \exp\left(-\frac{\pi}{4}
(-c_1 c_4 + c_2 c_3 + c_1 c_2 + c_3 c_4) \right) .
\end{equation}
We can readily verify that
\[ S^{\dagger}_{12} c_1 S_{12}= c_3 ,\hspace{1cm}
S^{\dagger}_{12} c_2 S_{12}= c_4
\]
i.e. $S_{12}$ swaps the roles of the pairs $(c_1,c_2)$ and
$(c_3,c_4)$. Similarly we have $S_{k,k+1}$ for any n.n. line pair
to swap pairs $(c_{2k-1}, c_{2k})$ and $(c_{2k+1}, c_{2k+2})$.
Note that the exponent in eq. (\ref{s12}) is n.n. quadratic so
$S_{12}$ is a n.n. Gaussian gate. In fact in the Jordan-Wigner
representation we get $S_{12}=(CZ)(SWAP)=G(Z,X)$.

Returning to $U_j$ and $H_j$, if $c_a$ and $c_b$ are not
associated to n.n. qubit lines we can use a ladder of $S_{k,k+1}$
conjugations to express $U_j$ as a product of at most $O(n)$ n.n.
$G(A,B)$'s. Thus starting from $U$ we obtain a product
$\tilde{U}=U_N \ldots U_1$ of at most $O(n^3)$ n.n. $G(A,B)$ gates
such that $V^\dagger c_\mu V= \tilde{U}^\dagger c_\mu \tilde{U}$
for all $c_\mu$. Hence this relation holds for all monomials
$c_{\mu_1}\ldots c_{\mu_k}$ too and thus for arbitrary matrices
$M$ (as the monomials span all, matrices) i.e. $V^\dagger M V=
\tilde{U}^\dagger M \tilde{U}$ for all $M$ so
$\tilde{U}=e^{i\delta} V$ for some overall phase $\delta$, which
may be set to zero by a further trivial $G(A,B)$ gate.\, $\Box$

We remark that theorem~\ref{gausscircuit} has a direct application
in digital quantum simulation (algorithmic quantum simulation by
the set of elementary gates) of a 1D quantum system whose
Hamiltonian $H$ is describable in the form of eq.~(\ref{quad_h}).
In particular, this includes the 1D XY Hamiltonian which exhibits
a quantum phase transition for suitable choice of its parameters.
We see that the real-time dynamics of the XY Hamiltonian {\em for
any length of time $t$} can be efficiently quantumly simulated in
terms of n.n. $G(A,B)$ gates. Another efficient circuit simulation
of the XY Hamiltonian was described recently in Verstraete {\em et
al.} 2008.

\section{Gaussian quantum circuits intertwined by Clifford
operations} \label{clops}

Recall (c.f. Nielsen and Chuang 2000) that the Pauli group $\cp_n$
on $n$ qubits contains all $n$-fold tensor products $P_1\otimes
\ldots \otimes P_n$ of Pauli matrices (i.e. each $P_j$ is $I,X,Y$
or $Z$) together with overall factors of $\pm 1$ and $\pm i$. An
$n$-qubit operation $T$ is a Clifford operation iff $T^\dagger A T
\in \cp_n$ for all $A\in \cp_n$ i.e. conjugation by $T$ preserves
the product Pauli structure. It is known (Gottesman 1997) that a
unitary operation $T$ is a Clifford operation iff it can be
expressed as a circuit of controlled-NOT (CNOT), Hadamard $H$ and 
$P={\rm diag}(1,i)$ gates.

With this in mind recall also that the Jordan-Wigner
representation of $\cc_{2n}$ comprises not only product operators,
but {\em Pauli} products. It is easy to verify that if a set of
hermitian operators $c_\mu$ satisfy the Clifford algebra relations
eq. (\ref{cliffcomm}) then so do $c'_\mu=V^\dagger c_\mu V$ for
any unitary $V$. Now recall that our classical simulation result
relied upon the quadratic Hamiltonian property in theorem
\ref{quadh} -- which in turn rests on the algebra relations eq.
(\ref{cliffcomm}) -- and the product structure of the matrix
representation (associated to product state inputs).  Hence if we
choose $V$ in $c'_\mu=V^\dagger c_\mu V$ to be a {\em Clifford
operation} $T$ we preserve {\em both} features and we can obtain
new classes of classically efficiently simulatable quantum
circuits using the Gaussian operations provided by the $c'_\mu$'s
 (assuming that the conditions of theorem \ref{simcost} are
also satisfied). Note that the Clifford operation $T$ itself
cannot generally be obtained as a Gaussian gate of the original
Clifford algebra representation $c_\mu$, nor thus by a circuit of
n.n. $G(A,B)$ gates.

The conjugation action of $T$ can be taken outside the quadratic
Hamiltonian and the exponential power series sum, showing that the
new Gaussian gate $U_{\rm new}= T^\dagger U_{\rm old} T$ is just
the original one $U_{\rm old}$ (e.g. n.n. $G(A,B)$'s) conjugated
by $T$. In a new circuit comprising new Gaussian gates $U_{\rm
new}$, the intermediate Clifford operations $T$ can be viewed as
cancelling each other by the unitarity identity $T T^{\dagger} =
I$. Thus we can alternatively think of these new simulatable
circuits as being the same as the old ones but the input states
are now $T\ket{\psi_{\rm in,old}}$ (now generally entangled) and
the final measurement is now $TZ_kT^\dagger$ (now generally a
multi-line observable) rather than $Z_k$ itself, i.e. we extend
the class of allowed inputs and measured outputs in theorem
\ref{one} while maintaining classical efficient simulatability.
From this point of view the new freedom associated to use of
Clifford operations $T$ appears at the {\em boundary} of circuits,
which is analogous to Valiant's use of basis changes in his
notion of holographic algorithm (Valiant 2007).

In our construction $T$ is generally a global ($n$-qubit)
operation, and we will require two further features:\\
(a) The Pauli operator $Z_k$ should be expressible as a {\em bounded}
degree polynomial in the $c'_\mu$'s. Recall that the classical simulation
cost depends on this degree $d$ as $O(n^d)$ (cf. theorem~\ref{simcost}),
and it was previously quadratic, but with arbitrary $T$'s we
may get $d=O(n)$.\\
(b) We wish to identify suitably {\em local} new gates $U_{\rm
new}$ acting on say, some constant number $K$ of qubit lines. For
general $T$ operators, even the conjugates of n.n. $G(A,B)$ gates
may become global $n$-qubit operators, so we may for example, seek
Clifford $T$'s such that these particular conjugates remain
$K$-local for some $K$. In contrast to (a) this requirement is not
essential for the existence of an efficient classical simulation
but it is desirable in view of the usual notion of quantum circuit
as comprising local gates each acting on a bounded number of
lines.

We also remark that, in the above construction, we need to choose
a Clifford operation $T_n$ for each number $n$ of qubit lines. A
curious feature is that in addition to being able to vary the
structure of $T_n$ with $n$, each $T_n$ need not itself be
``translationally uniform'' across the $n$ lines whereas the class
of all n.n. $G(A,B)$ gates as a whole does have a translationally
uniform structure. Hence we can obtain classically simulatable
quantum circuits which have different kinds of gates allowed on
different sections of the qubit line set.

\begin{figure}[t]
\begin{center}
\includegraphics[width=0.6\textwidth,clip]{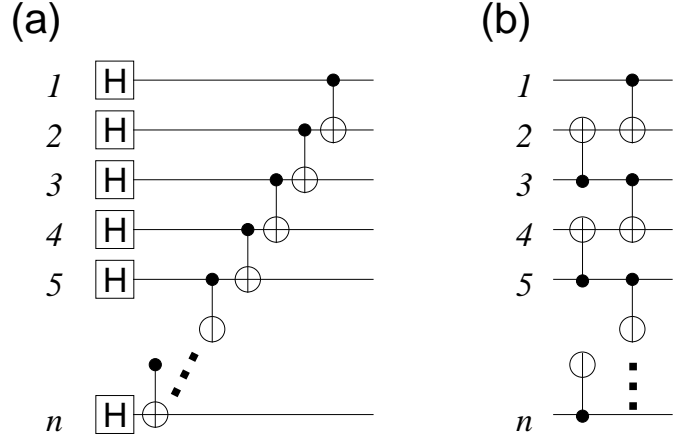}
\end{center}
\caption{Intertwining Clifford operations $T$ for the examples 2
and 3 in (a) and (b) respectively, in order to change the
representation of the Clifford algebra.} \label{fig:intertwineT}
\end{figure}

To conclude this section we give three illustrative examples of
this construction.

\noindent {\bf Example 1.} \,\, Clearly any circuit $T$ of $SWAP$
operations is a Clifford operation. In this case $T^\dagger
G(A,B)_{\rm n.n.}T$ amounts to allowing the $G(A,B)$ gates to act
on correspondingly selected distant lines. However in any such
resulting Gaussian circuit the lines may always be simply
re-ordered to restore all $G(A,B)$'s to n.n. status.

\noindent {\bf Example 2.}\,\, Let $CNOT_{i \,j}$ denote the
$n$-qubit operation that applies the 2-qubit $CNOT$ gate
with control line $i$ and target line $j$. Let $H_i$ denote the
1-qubit Hadamard gate on line $i$. Consider the (translationally
uniform) Clifford operation:
\[ T_n = CNOT_{1 \,2}CNOT_{2 \,3} \ldots CNOT_{n-1 \, n} H_1 H_2
\ldots H_n \] as depicted in figure \ref{fig:intertwineT} (a).
Indeed, this $T$ operation is known as a duality transformation
of a 1D quantum system (cf. Plenio 2007). Conjugating the
Jordan-Wigner representation $c_\mu$'s of eq.~(\ref{jwr}), we
obtain $c'_\mu = T^\dagger c_\mu T$ giving:
\[ c'_{2k-1}=X_{k-1} (\prod_{j=k}^n Z_j ) \hspace{1cm}
c'_{2k} = -Y_k (\prod_{j=k+1}^n Z_j) \] so that
$Z_k=-i c'_{2k} c'_{2k+1}$ remains quadratic in the
generators.
The six n.n. quadratic Hamiltonian terms on lines $k,k+1$ are
\[
 i\left(\alpha_0 c_{2k-1}c_{2k+2} - \alpha_1 c_{2k}c_{2k+1} +
\beta_1 c_{2k-1}c_{2k+1} - \beta_2 c_{2k}c_{2k+2} - \gamma_1
c_{2k-1}c_{2k} - \gamma_2 c_{2k+1}c_{2k+2}\right) \]
 and correspond in the Jordan-Wigner representation to
 \[ \alpha_{0} Y_k Y_{k+1} + \alpha_{1} X_k X_{k+1}
 + \beta_{1} Y_k X_{k+1} + \beta_{2} X_k Y_{k+1}
 + \gamma_{1} Z_k + \gamma_{2} Z_{k+1}.
\]
Conjugation by $T$ gives the Hamiltonian (also known as the
three-body cluster-state interaction):
\[
- \alpha_{0} X_{k-1}Z_{k}X_{k+1} - \beta_{1} X_{k-1} Y_{k} -
\beta_{2} Y_{k} X_{k+1} + \gamma_{1} X_{k-1} X_{k} + \gamma_{2}
X_{k}X_{k+1} + \alpha_{1} Z_{k}.
\]
Note that this Hamiltonian (although quadratic in the
$c'_\mu$'s) has now become 3-local so that the 6 parameter
family of n.n. $G(A,B)$'s on lines $k,k+1$ will conjugate to a 6
parameter family of 3-local gates on lines $k-1,k,k+1$ (and we omit
computation of the explicit form of these 3-qubit gates). Since we
have expanded into 3 lines we may go back and consider arbitrary
quadratic Hamiltonian terms in the $c_\mu$'s associated to lines
$k-1,k,k+1$, involving $^6 C_2 = 15$ parameters. By computing their
conjugates under $T$ we find that they all remain 3-local, giving
a 15 parameter family of 3-local Gaussian gates. However by
theorem \ref{gausscircuit}, any member of this 15 parameter family
is obtainable as a circuit of the initial 6 parameter family.
Finally, we see, by the construction, that arbitrary poly-sized
circuits of the 15 parameter family of 3-local gates, with input
product states and a final $Z_k$ measurement can be classically
efficiently simulated.

\noindent {\bf Example 3.}\,\, For odd $n$ consider the
translationally uniform Clifford operation $T_n$ given by
\[ T_n = (CNOT_{1 \,2}CNOT_{3 \,4} \ldots CNOT_{n-2 \, n-1})
(CNOT_{3 \, 2}CNOT_{5 \,4}\ldots CNOT_{n \, n-1}) \] as
depicted in figure \ref{fig:intertwineT} (b). Conjugating the
Jordan-Wigner representation we obtain in this case:
\[
\begin{array}{rcl}
c'_{4l} & = &\left(\prod_{j=2}^{2l-1} Z_{j}\right) Y_{2l} Z_{2l+1} , \\
c'_{4l+1} & = &\left(\prod_{j=2}^{2l-1} Z_{j}\right) Y_{2l} Y_{2l+1}
X_{2l+2}, \\
c'_{4l+2} & = &\left(\prod_{j=2}^{2l-1} Z_{j}\right) Y_{2l} X_{2l+1}
X_{2l+2}, \\
c'_{4l+3} & = &\left(\prod_{j=2}^{2l} Z_{j}\right) X_{2l+2},
\end{array}
\]
supplemented by boundary terms
\[
c'_1 = X_{1} X_{2} \hspace{1cm} c'_2 = Y_{1} X_{2}\hspace{1cm} c'_3 =
Z_{1} X_{2}.
\]

It follows from these expressions that the conjugations of n.n.
$G(A,B)$ gates on lines $k,k+1$ become 4-local gates on lines
$k,k+1,k+2,k+3$. By considering all possible quadratic terms of
$c_\mu$'s associated to these 4 lines we obtain for each $k$, a 13
parameter family of Gaussian gates (i.e. not all $^8C_2$ quadratic
terms remain 4-local under conjugation). These are generated by
the following Hamiltonians and their commutators:
\[ \begin{array}{rl}
\mbox{for $k$ odd:} &  Z_{k}Z_{k+1}X_{k+2}X_{k+3},\,\,
Z_{k}Z_{k+1}Z_{k+2},\,\, X_{k+1}Z_{k+2}X_{k+3},\,\, X_{k}X_{k+1}, \\
 & X_{k+1}X_{k+2},\,\, X_{k+2}X_{k+3},\,\, Z_{k},\,\, Z_{k+2};\\ & \\
\mbox{for $k$ even:} &  X_{k}X_{k+1}Z_{k+2}Z_{k+3},\,\,
Z_{k+1}Z_{k+2}Z_{k+3},\,\, X_{k}Z_{k+1}X_{k+2},\,\, X_{k}X_{k+1},\\
& X_{k+1}X_{k+2},\,\, X_{k+2}X_{k+3},\,\, Z_{k+1},\,\, Z_{k+3} .
\end{array}
\]
Thus when $k$ is odd $Z_k$ is obtained as a quadratic expression
in the $c'_\mu$'s, whereas when $k$ is even $Z_k$ requires a sixth
degree Clifford algebra monomial viz. the product of of
$Z_{k-1}Z_{k}Z_{k+1}$, $Z_{k-1}$ and $Z_{k+1}$ each of which is in
the $k$-even list above and hence quadratically representable.
Thus arbitrary poly-sized circuits of the 26 parameter family of
4-qubit gates defined by the Hamiltonians above, are classically
efficiently simulatable albeit with a higher simulation cost which
now scales as $O(n^6)$.

\section{Concluding remarks}\label{concls}

In theorems \ref{one} and \ref{two} we have seen that quantum
computational power may be made to appear as a surprisingly
delicate extension of its classical counterpart. Is it conceivable
that the passage from n.n. to next-n.n. use of $G(A,B)$ gates may
be achieved while maintaining classical simulatability? We relate
this question to some more formal complexity theoretic
considerations after introducing some further terminology.

Recall that BQP is the class of languages decided by a uniform
(poly-sized) family of quantum circuits for which, given any input
computational basis state, each output probability $p_0$ and $p_1$
is $\geq \frac{2}{3}$ or $\leq \frac{1}{3}$ (with output 1 resp. 0
designating acceptance resp. rejection of the input). Introduce
PQP (a quantum analogue of the classical class PP, c.f.
Papadimitriou 1994) to denote the corresponding class of languages
for which the bounded probability conditions are relaxed to
requiring only that $p_0$ and $p_1$ are $\geq
\frac{1}{2}+\frac{1}{2^n}$ or $ \leq \frac{1}{2}-\frac{1}{2^n}$.
Clearly BQP $\subseteq$ PQP but we also have NP $\subseteq$ PQP
(e.g., using a quantum algorithm for SAT that simply computes a Boolean
function on an equal superposition of all its inputs and measures
the function output register for values 0 versus 1). Similarly it
is straightforward to see that PP $\subseteq$ PQP but furthermore
it may be shown (Watrous 2008) that PP $=$ PQP.

Now let V$_{{\rm n.n.}} \subseteq $ PQP be the class of languages
decided by PQP-circuits of n.n. $G(A,B)$ gates. With our strong
notion of classical simulation, theorem \ref{one} gives V$_{\rm
n.n.} \subseteq $ P. Also theorem \ref{two} shows that every
language in PQP is decidable (relative to the PQP probability
conditions) by a circuit comprising only n.n. and next-n.n.
$G(A,B)$ gates (and applied to a suitably restricted set of
inputs, encoding strings of 0's and 1's). Thus if the latter were
also classically simulatable we would have P $=$ NP $=$ PP i.e. in
the context of the PQP probability conditions, an extra
supra-classical computational power {\em must} be associated to
the single distance extension of the range of n.n. 2-qubit
$G(A,B)$ gates if these classical computational complexity classes
are to be unequal.

On the other hand the same analysis carried out relative to the
(far more stringent) BQP probability conditions (viz. requiring
$p_0$ and $p_1$ to be bounded away from $\frac{1}{2}$ by at least
$\frac{1}{6}$) is less compelling. Indeed it is generally believed
(although not proven) that neither NP nor PP is contained in BQP
so in the context of BQP circuits it becomes less implausible that
the passage from n.n. to next n.n. $G(A,B)$ circuits might retain
classical simulatability (now no longer implying equality of P, NP
and PP). But then we would have P $=$ BQP. Actually, more simply,
to obtain BPP $=$ BQP it would suffice to simultaneously relax our
(very strong) notion of classical simulation to a far weaker
requirement viz. the ability to merely {\em sample} the output
distribution {\em once} by classical efficient means, in contrast
to classically efficiently computing the probabilities to
exponential accuracy.

\noindent {\large\bf Acknowledgments} RJ and AM acknowledge
support from the EC network QICS which provided collaborative
opportunities for this work. AM acknowledges discussions with H.J.
Briegel, B. Kraus, and R. Somma, and is supported by FWF and the
EC networks OLAQUI, SCALA. RJ is supported by EPSRC QIP-IRC and EC
network QAP.
\bigskip

\noindent {\Large\bf References}
\bigskip

\setlength{\parindent}{0mm}

Bravyi, S. 2005 Lagrangian representation for fermionic linear
optics. {\em Quant. Inf. Comp.} {\bf 5}, 216-238.

Bravyi, S. 2008 Contraction of matchgate tensor networks on
nonplanar graphs. arXiv:0801.2989v1.

Bravyi, S. and Kitaev, A. 2002 Fermionic quantum computation. {\em
Annals of Physics} {\bf 298:1}, 210-226.

Cai, J-Y. and  Choudhary, V. 2006a Valiant's holant theorem and
matchgate tensors. {\em Lecture Notes in Computer Science} {\bf
3959}, 248-261.

Cai, J-Y. and Choudhary, V. 2006b On the theory of matchgate
computations. ECCC TR06-018.

Cleve, R. and  Watrous, J. 2000  Fast parallel circuits for the
quantum fourier transform. {\em Proc. 41st Annual Symposium on
Foundations of Computer Science}, 526-536.

Deutsch, D. and Jozsa, R. 1992 Rapid solution of problems by
quantum computation.  {\em Proc. R. Soc. Lond. A} {\bf 439}.
553-558.

DiVincenzo, D. and Terhal, B. 2005 Fermionic linear optics
revisited. {\em Found. Phys.} {\bf 35}, 1967-1984.

Gottesman, D. 1997 {\em Stabilizer Codes and Quantum Error
Correction}, PhD thesis, California Institute of Technology,
Pasadena, CA.

Hoffman, D., Raffenetti, R. and Ruedenberg, K. 1972 Generalisation of
Euler angles to N dimensional orthogonal matrices. {\em J. Math.
Phys.} {\bf 13}, 528-532.

Jerrum, M. 2003 {\em Counting, Sampling and Integrating:
Algorithms and Complexity}. Birkhauser, Basel, Switzerland.

Jordan, P. and Wigner, E. 1928 \"{U}ber das Paulische
\"{A}quivalenzverbot. {\em Zeitschrift f\"{u}r Physik} {\bf 47},
631-651.

Jozsa, R. 2006 On the simulation of quantum circuits.
arXiv:quant-ph/0603163.

Kasteleyn, P. 1961 The statistics of dimers on a lattice. {\em
Physica} {\bf 27}, 1209-1225.

Kempe, J., Bacon, D., DiVincenzo, D. and Whaley, K. 2001 Encoded
universality from a single physical interaction. {\em Quant. Inf.
Comp.} {\bf 1} (Special issue December 2001) 33-55.

Knill, E. 2001 Fermionic linear optics and matchgates.
arXiv:quant-ph/0108033.

Markov, I. and Shi, Y. 2008 Simulating quantum computation by
contracting tensor networks. {\em SIAM J. Computing} {\bf 38},
963-981.

Nielsen, M. and Chuang, I.  2000 {\em Quantum Computation and
Quantum Information}. Cambridge University Press.

Papadimitriou, C. 1994 {\em Computational Complexity}.
Addison-Wesley, Reading, MA.

Plenio, M.B., 2007 Remarks on duality transformations and
generalised stabiliser states. {\em J. Mod. Optics}, {\bf 54},
2193-2201.

Raz, R. 1999 Exponential separation of quantum and classical
communication complexity. {\em Proc 31st Annual ACM Symp. Theory
of Computing (New York: ACM Press)}, 358-367.

Shor, P. 1997 Polynomial time algorithms for prime factorisation
and discrete logarithms on a quantum computer. {\em SIAM J.
Computing} {\bf 26}, 1484-1509.

Simon, D. 1997 On the power of quantum computation. {\em SIAM J.
Computing} {\bf 26}, 1474-1483.

Somma, R., Barnum, H., Ortiz, G., and Knill, E. 2006 Efficient
solvability of hamiltonians and limits on the power of some
quantum computational models. {\em Phys. Rev. Lett.} {\bf 97},
190501.

Temperley, H. and Fisher, M. 1961 M. E. dimer problems in
statistical mechanics -- an exact result. {\em Philosophical
Magazine} {\bf 6}, 1061-1063.

Terhal, B. and  DiVincenzo, D. 2002 Classical simulation of
nonintercating-fermion quantum circuits. {\em Phys. Rev. A} {\bf
65}, 032325/1-10.

Terhal, B. and  DiVincenzo, D. 2004 Adaptive quantum computation,
constant depth circuits and Arthur-Merlin games. {\em Quant. Inf.
Comp.} {\bf 4(2)}, 134-145.

Valiant, L. 2002 Quantum circuits that can be simulated classically
in polynomial time. {\em SIAM J. Computing} {\bf 31:4}, 1229-1254.

Valiant, L. 2007  Holographic algorithms. {\em SIAM J. Computing}
{\bf 37:5} 1565-1594.

Verstraete, F., Cirac, J.I., and Latorre, J.I. 2008 Quantum
circuits for strongly correlated quantum systems. arXiv:0804.1888.

Watrous, J. 2008 Quantum computational complexity.
arXiv:0804:3401v1. (To appear in Springer Encyclopedia of
Complexity and Systems Sciences).

Yoran, N. and  Short, A. 2006 Classical simulation of
limited-width cluster-state quantum computation. {\em Phys. Rev.
Lett.} {\bf 96}, 170503.

\end{document}